\documentclass[preprint,amsmath,amssymb,superscriptaddress,floatfix]{revtex4-2}
\usepackage{graphicx}
\usepackage{makecell}
\usepackage{multirow}
\usepackage{booktabs}
\usepackage{longtable}
\usepackage{lineno} 
\usepackage{color}
\usepackage{hyperref}

\begin{document}

% \linenumbers

\title{Finite-temperature properties of antiferroelectric perovskite $\rm PbZrO_3$ from deep learning interatomic potential}
\author{Huazhang Zhang}
\email[Corresponding author: ]{hzhang@uliege.be}
\address{Theoretical Materials Physics, Q-MAT, University of Liège, B-4000 Sart-Tilman, Belgium}
\address{Department of Physics, School of Science, Wuhan University of Technology, Wuhan 430070, People’s Republic of China}
\author{Hao-Cheng Thong}
\address{Theoretical Materials Physics, Q-MAT, University of Liège, B-4000 Sart-Tilman, Belgium}
\address{State Key Laboratory of New Ceramics and Fine Processing, School of Materials Science and Engineering, Tsinghua University, Beijing 100084, People's Republic of China}
\author{Louis Bastogne}
\address{Theoretical Materials Physics, Q-MAT, University of Liège, B-4000 Sart-Tilman, Belgium}
\author{Churen Gui}
\address{Theoretical Materials Physics, Q-MAT, University of Liège, B-4000 Sart-Tilman, Belgium}
\address{Key Laboratory of Quantum Materials and Devices of Ministry of Education, School of Physics, Southeast University, Nanjing 211189, People’s Republic of China}
\author{Xu He}
\address{Theoretical Materials Physics, Q-MAT, University of Liège, B-4000 Sart-Tilman, Belgium}
\author{Philippe Ghosez}
\email[Corresponding author: ]{philippe.ghosez@uliege.be}
\address{Theoretical Materials Physics, Q-MAT, University of Liège, B-4000 Sart-Tilman, Belgium}
\date{\today}

\begin{abstract}

The prototypical antiferroelectric perovskite $\rm PbZrO_3$ (PZO) has garnered considerable attentions in recent years due to its significance in technological applications and fundamental research.
Many unresolved issues in PZO are associated with large length- and time-scales, as well as finite temperatures, presenting significant challenges for first-principles density functional theory studies.
Here, we introduce a deep learning interatomic potential of PZO, enabling investigation of finite-temperature properties through large-scale atomistic simulations. 
Trained using an elaborately designed dataset, the model successfully reproduces a large number of phases, in particular, the recently discovered 80-atom antiferroelectric $Pnam$ phase and ferrielectric $Ima2$ phase, providing precise predictions for their structural and dynamical properties.
Using this model, we investigated phase transitions of multiple phases, including $Pbam$/$Pnam$, $Ima2$ and $R3c$, which show high similarity to the experimental observation.
Our simulation results also highlight the crucial role of free-energy in determining the low-temperature phase of PZO, reconciling the apparent contradiction: $Pbam$ is the most commonly observed phase in experiments, while theoretical calculations predict other phases exhibiting even lower energy.
Furthermore, in the temperature range where the $Pbam$ phase is thermodynamically stable, typical double polarization hysteresis loops for antiferroelectrics were obtained, along with a detailed elucidation of the structural evolution during the electric-field induced transitions between the non-polar $Pbam$ and polar $R3c$ phases.

\end{abstract}

\maketitle

\newpage

\section{Introduction}
Antiferroelectric (AFE) materials refer to a class of crystals that typically possesses local antiparallel dipoles, which forms a macroscopic non-polar state, but can be turned into a polar state under the application of an electric field \cite{RabeAFE, LinesGlass}.
The electric-field induced non-polar to polar switching gives rise to the peculiar double polarization-versus-electric field ($P$-$E$) hysteresis loop, which provides antiferroelectrics a series of functional properties that are highly appealing for applications \cite{RN408, RN536, RN532, RN526, RN903}.
Lead zirconate ($\rm PbZrO_3$, PZO) was historically the first discovered antiferroelectric material, right after the concept proposed by Kittel in the early 1950s \cite{KittelAFE, RN533, RN534}.
PZO has received considerable attention since then, primarily driven by the strong demand of  applications, including energy storage, electromechanical actuation, electrocaloric effects, thermal switching, etc \cite{RN724, RN936, RN977, RN983, RN1017, RN407, ECE2006science, RN902}.

\begin{figure*}[h]
\centering
\includegraphics[scale=0.5]{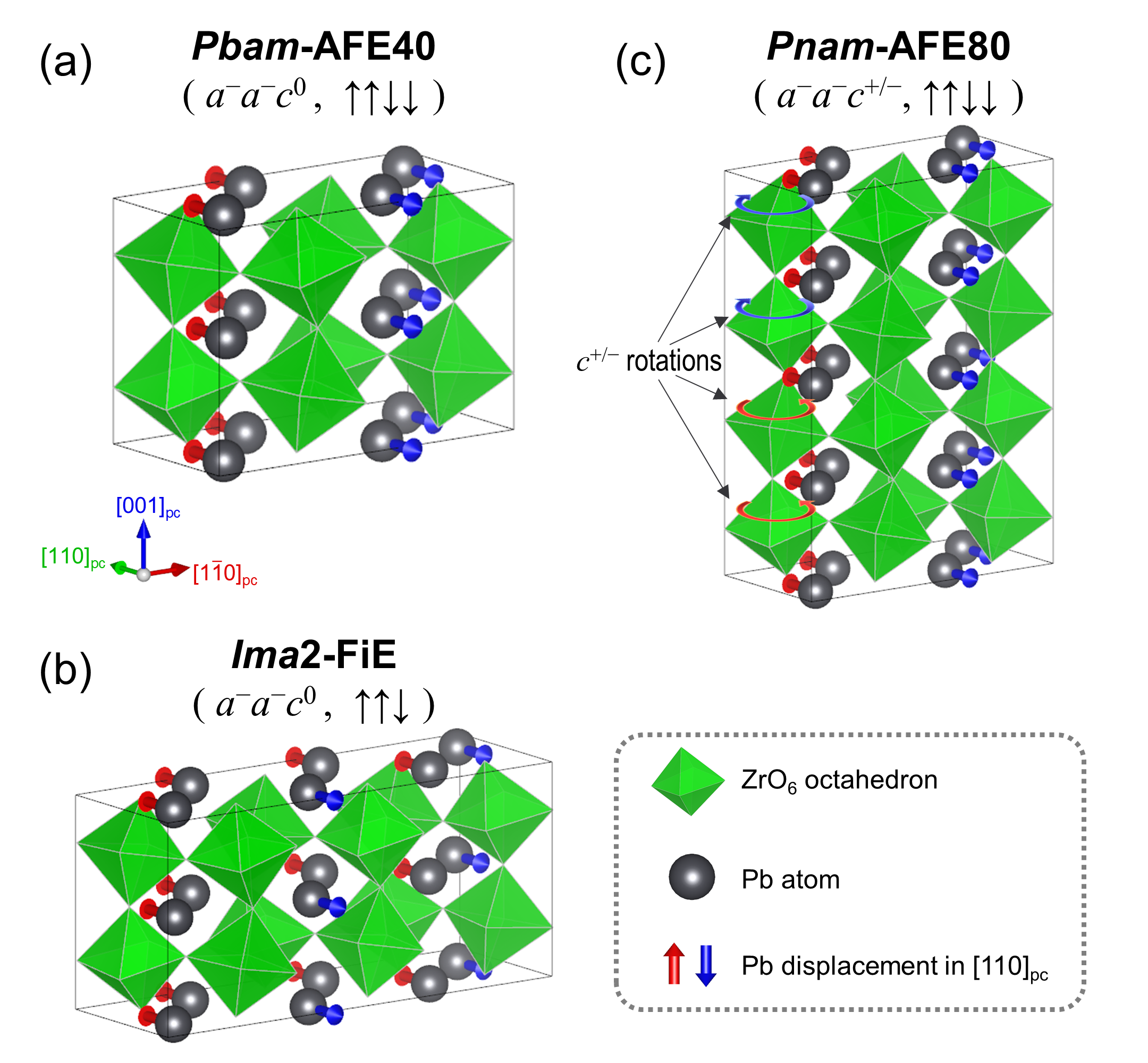}
\caption{Schematic views of the structures of (a) $Pbam$-AFE40, (b) $Ima2$-FiE, and (c) $Pnam$-AFE80 phases of PZO. 
The $Pbam$-AFE40 phase is featured by an octahedra rotation pattern $a^-a^-c^0$ and a fourfold periodic Pb displacement pattern ``$\uparrow \uparrow \downarrow \downarrow$'' in the pseudocubic $[110]$ direction. 
The $Ima2$-FiE phase has a octahedra rotation pattern $a^-a^-c^0$ similar to that of the $Pbam$-AFE40 phase, but with a different threefold periodic Pb displacement pattern ``$\uparrow \uparrow \downarrow$'' in the pseudocubic [110] direction. 
In the $Pnam$-AFE80 phase, the Pb displacement pattern is similar to that of the $Pbam$-AFE40 phase, but there are additional octahedra rotations around the $c$-axis (pseudocubic $[001]$), alternating between in-phase and anti-phase rotations, i.e., two layers of clockwise rotations followed by two layers of anticlockwise rotations, and so on. 
We refer to this alternating in-phase and anti-phase $c$-rotation pattern as ``$c^{+/-}$''.}
\label{fig:struct}
\end{figure*}

However, many fundamental properties of PZO have remained not fully understood, particularly those related to large length- and time-scales and finite temperatures, e.g., the true ground state and the temperature-dependent phase diagram.
According to the experimental observations, the ground state of PZO was found to be a 40-atom antiferroelectric $Pbam$ phase [$Pbam$-AFE40, Fig. \ref{fig:struct}(a)] \cite{PZO_Pbam}, with a transition to the high-temperature cubic phase occurring at $T_{\rm C} \approx 505$ K \cite{RN533}. 
Some studies have also suggested a controversial intermediate phase within a narrow temperature range around $T_{\rm C}$ \cite{RN907, RN687, RN688, RN128, RN50, RN144}. 
Recently, the traditional viewpoint regarding the $Pbam$ ground state of PZO has even been challenged by theoretical discoveries of a ferrielectric $Ima2$ phase [$Ima2$-FiE, Fig. \ref{fig:struct}(b)] \cite{PZO_30GS_Ima2}, and an 80-atom antiferroelectric $Pnam$ phase [$Pnam$-AFE80, Fig. \ref{fig:struct}(c)] \cite{PZO_80GS_Pnam}, which were found to exhibit lower energies than that of the $Pbam$ phase. 
The physical properties of these newly predicted phases and their stability against temperature are yet to be clarified.
While recent experimental observations have revealed the existence of $Ima2$-like structures locally in PZO \cite{RN853, RN933}, and the regions of which could be further coarsened under compressive strains \cite{RN999}, the possibility of the existence of the $Ima2$-FiE phase at a macroscopic scale remains unclear. 
This raises a specific question: how to reconcile the apparent contradiction between these two aspects, where theoretical predictions suggest a lower energy for $Ima2$-FiE, but experimental observations more commonly identify $Pbam$-AFE40 as the low-temperature structure of PZO?

Another aspect yet to be fully understood in PZO  concerns its electric-field induced nonpolar-polar transitions, which constitute the conceptual foundation for PZO being AFE and enable most of the functionalities of PZO-based materials.
From theoretical calculation, polar $R3c$ and $Ima2$-FiE are found to be energetically very close to the non-polar antiferroelectric $Pbam$-AFE40 phase (with energy differences around 1 meV/f.u. \cite{RN101, RN308, PZO_30GS_Ima2}), and thus become possible polar phases induced by electric field.
Experimentally, however, a non-polar to polar phase transition in PZO bulk samples (crystal or ceramic) is hardly reported since the electric field required is usually higher than the breakdown field \cite{RN533, RN823}. 
Thus, most experimental studies on the electric-field induced phase transitions were conducted on PZO films \cite{RN717, RN817, RN702} or PZO-based solid solutions \cite{RN803, RN429, RN816, RN426, RN409, RN135, RN919}.
However, the influences from strains, surfaces and foreign elements in films and solid solutions are usually significant and cannot be ignored.
Given the fundamental importance of nonpolar-polar transitions in pure PZO and the experimental challenges with bulk materials, it is necessary and potentially feasible to study the electric-field induced transitions employing reliable theoretical methods.
The outcomes would also be beneficial to the understanding and optimization of the performance of PZO-based materials and facilitate their applications.

The aforementioned issues in PZO could be resolved through theoretical investigations relying on a large configuration system, and its dynamical response to the external fields such as temperature and electric field. 
Although first-principles density function theory (DFT) calculation has been a successful atomistic simulation method for investigating PZO, it is computationally expensive for such dynamical simulation, especially for a system with more than hundreds of atoms. 
Molecular dynamics (MD) or Monte-Carlo simulations using first-principles-based effective interatomic potentials have become an appealing solution, i.e. following the so-called ``second-principles'' approach \cite{RN23}.
Relevant efforts include the development of effective Hamiltonian and shell models for PZO \cite{RN52, RN703, RN962, RN136}. 
However, the investigation of a system like PZO with a complex energy surface remains challenging, resulting poor prediction of certain crucial structures. 

In recent years, machine learning methods have been introduced for constructing interatomic potentials, offering new solutions for highly accurate large-scale simulations that consider all atomistic degrees of freedom \cite{RN992, RN967}.
Particularly, the deep potential for molecular dynamics (DeePMD) has attracted significant attention \cite{RN963}. 
Thanks to the development of the open-source DeePMD-kit software, this method is gaining increasing popularity in studies across a wide variety of systems, including water, metals, oxides, organic molecules, etc \cite{RN753, RN993, RN996, RN997}.
For ferroelectric compounds, the deep potential approach has been successfully applied to $\rm SrTiO_3$ \cite{RN1}, $\rm KNbO_3$ \cite{RN586}, $\rm PbTiO_3$ \cite{RN167}, $\rm Pb(Zr, Ti)O_3$ \cite{RN1013}, $\rm HfO_2$ \cite{RN9}, $\rm PbTe$ \cite{RN969}, $\rm In_2Se_3$ \cite{RN10}, bi-layered $\rm BN$ \cite{RN991}, etc.
Recently, a modular development scheme for deep potential in complex solid solutions was proposed \cite{RN768}, and a universal model for perovskite oxides involving 14 metal elements was presented \cite{RN966}.
Regarding the antiferroelectric PZO, despite its highly complex potential energy surface, given the success of previous applications, employing the deep potential method appears very promising for developing a highly accurate interatomic potential.

In this work, we developed a deep learning interatomic potential of PZO for finite-temperature simulation. 
The model, which is trained with an elaborately designed training set, can accurately reproduce numerous phases in PZO and provide precise predictions on their properties. 
Notably, the model successfully captures the recently discovered $Pnam$-AFE80 and $Ima2$-FiE phases.
The model is thoroughly validated and utilized for investigating the finite-temperature properties through molecular dynamic simulations.
The simulations successfully reproduce the temperature-dependent phase transitions and double $P$-$E$ loops, showing good agreement with experiments.
Additionally, the simulations also provide insights into why PZO tends to adopt the $Pbam$-AFE40 as its room-temperature phase, rather than other low energy phases like $Ima2$-FiE.
We prospect that this model will serve as a useful tool for studying PZO, offering opportunities for new insights into this fascinating material.

\section{Methodology}

\subsection{First-principles calculations}
The first-principles DFT calculations were performed using the \textsc{Abinit} software \cite{Gonze2020, Gonze2016, Gonze2009, Gonze2002}, employing the GGA-PBEsol functional \cite{PBEsol} and a planewave-pseudopotential approach with optimized norm-conserving pseudopotentials from the PseudoDojo server  \cite{VANSETTEN201839, Hamann2013}.
The energy cutoff for the plane-wave expansion was 60 Ha, and the Brillouin zone sampling was equivalent or denser to a $6 \times 6 \times 6$ $k$-point grid for the 5-atom perovskite unit cell.
The electronic self-consistent cycles were converged until the potential residual is smaller than $10^{-18}$ $\rm Ha$.
Structural relaxations were performed based on the Broyden-Fletcher-Goldfarb-Shanno minimization algorithm until the forces are less than $10^{-6}$ $\rm Ha/Bohr$ and the stresses are less than $10^{-8}$ $\rm Ha/Bohr^3$.
The Born effective charges, phonon dispersions, and elastic and piezoelectric tensors were calculated according to the density functional perturbation theory (DFPT) as implemented in \textsc{Abinit} and analyzed using the \textsc{Anaddb} tool \cite{DFPT}.

For building the training set, we used a less stringent convergence criterion for the electronic self-consistent cycles of $10^{-10}$ $\rm Ha$ on the potential residual to facilitate the acquisition of the data.
We also developed a plugin for the dpdata package to convert the data format \cite{dpdata_abinit}, so that the DeePMD-kit can make use of the \textsc{Abinit} results to train models.

\subsection{Second-principles calculations}
\subsubsection{Model training}
The deep learning interatomic potential model for PZO was trained using the open-source package DeePMD-kit \cite{DPMD1, DPMD2}. 
The deep potential method assumes that the total energy of a system is the summation of the energy contributions from each of the atoms, whereas the energy contribution from each atom is determined by its local environment within a cutoff radius and can be parameterized with neural networks.
To build the PZO model, the cutoff radius was set to be 9.0 \AA, smoothing from 1.5 \AA.
The embedding and fitting networks both have three layers, with the size of $(25, 50, 100)$ and $(240, 240, 240)$, respectively.
The loss function is defined by
\begin{equation}
\begin{split}
L = p_e (\Delta e)^2 + p_f \frac{\sum_i{\mid\Delta f_i\mid^2}}{3N} +p_\xi \frac{\parallel\Delta \xi \parallel^2}{9}
\end{split},
\label{eq:loss}
\end{equation}
where $\Delta$ denotes the difference between the model predictions and the training data, $e$ is the energy per atom, $f_i$ is the force vector on the atom $i$ and $N$ is the number of atoms, $\xi$ is the virial.
The $p_e$, $p_f$ and $p_\xi$ are the prefactors which balance the losses in energy, forces and virials, respectively. 
The $p_e$ and $p_\xi$ increase from 0.02 to 1, and $p_f$ decreases from 1000 to 1 during the training procedure.

\subsubsection{Model-based calculations}
The model-based calculations were carried out using \textsc{Lammps} \cite{lammps1, lammps2} with periodic boundary conditions.
Structural relaxations were performed based on the conjugate gradient algorithm, with the convergence criterion on forces of $5 \times 10^{-3}$ $\rm eV/\AA$.
Phonon dispersion curves were calculated based on the finite difference method, using \textsc{Phonopy} \cite{phonopy} to analyse the force set produced by the model and extracted by \textsc{Lammps}.
The non-analytical corrections for the phonon dispersions were conducted using the Born effective charges of the corresponding phases from DFPT calculations.
The elastic and piezoelectric tensors were calculated by analyzing the stress and polarization changes to strain perturbations, respectively, with the atomic positions fully relaxed.
The finite-temperature MD simulations were performed using NPT ensemble, with the temperature and pressure controlled by a Nose-Hoover thermostat and barostat. 
The electric field was applied by adding  on each of the atoms extra forces according to $f_{i,\alpha} = \sum_{\beta} Z^{\ast}_{i,\alpha,\beta}E_{\beta}$, where $Z^{\ast}$ is the Born effective charges in the cubic reference phase, $E$ is the applied electric field, the subscript $i$ is the index of atoms, and $\alpha$, $\beta$ denote the Cartesian directions $x$, $y$ and $z$.
The time step for the MD simulations is 1 fs. 
At each temperature, the system was first equilibrated for 50 ps, then the simulation continued for another 50 ps for property analysis.
The MD trajectories were analyzed with \textsc{Agate} \cite{agate}.
In particular, the polarization was calculated based on the Born effective charge of the cubic phase using the algorithm implemented in \textsc{Agate}.

\section{Second-principles model}
\subsection{Training set design}

\begin{figure*}[htb]
\centering
\includegraphics[scale=0.5]{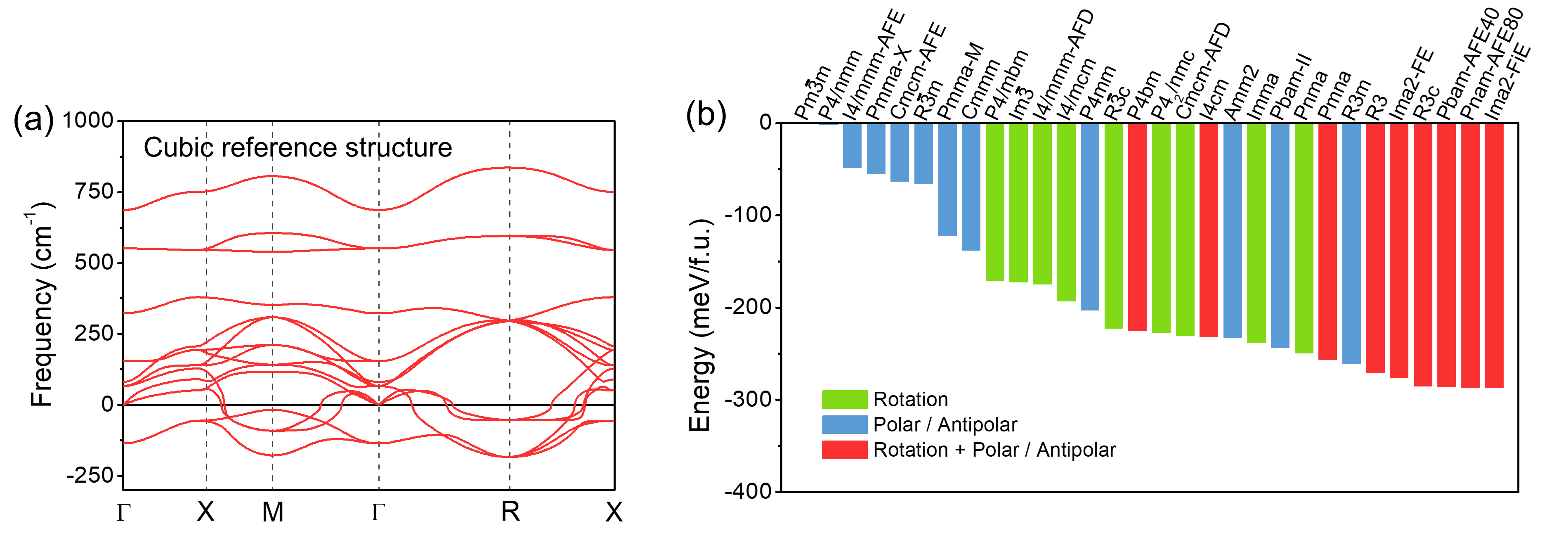}
\caption{Exploration of potential energy surface of PZO by first-principles calculations.
(a) Phonon dispersions of PZO in the cubic reference structure.
(b) Energies of various stationary phases of PZO, where the cubic phase is taken as the energy reference.}
\label{fig:pes}
\end{figure*}

PZO probably has one of the most complex potential energy surfaces of all perovskite oxides.
As can be seen from the phonon dispersion curves of its cubic parent phase [Fig. \ref{fig:pes}(a)], lattice instabilities are present throughout the whole Brillouin zone.
The strongest instabilities include the polar modes at $\Gamma$, and the in-phase and anti-phase oxygen octahedra rotations at $M$ and $R$.
Additionally, there are also unstable modes at $X$, $M$ and $R$ points, which are featured by the antipolar motions of cations.
By successively condensing unstable modes in the cubic parent phase, PZO can reach various stationary phases.
Fig. \ref{fig:pes}(b) and Table \ref{tab:phases} provide a series of stationary phases of PZO, where the cubic parent phase is taken as the energy reference.
It is worth noticing that not only the number of different phases in PZO is considerably large, but also many of them are very close in energy. 
So, it seems quite challenging to capture such a large number of competing phases within a single interatomic effective potential model.

\begin{figure*}[htb]
\centering
\includegraphics[scale=0.5]{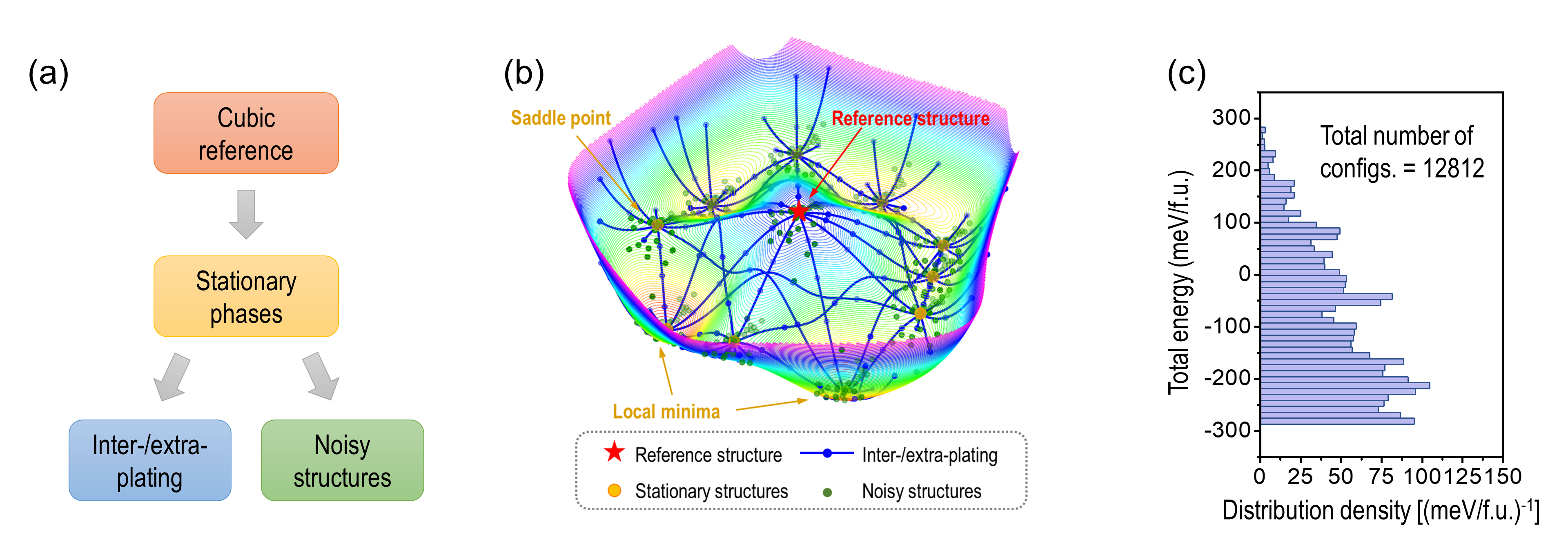}
\caption{Training set design.
(a) Procedures of sampling the potential energy surface and constructing the training set.
(b) Schematic plot of sampling the potential energy surface.
(c) Energy distribution of the configurations in the training set.}
\label{fig:trainingset}
\end{figure*}

\begin{table*}[b]
\tiny
\caption{Space group, lattice distortions and calculated energies of various phases of PZO. 
For the lattice distortions, only the dominant antiferrodistortive (AFD) rotations and polar/antipolar cation motions are presented.
The DFT and model energies are reported in relative values with respect to the cubic reference structure.
}
\label{tab:phases}
\begin{ruledtabular}
\begin{tabular}{ccccccc}
 & Phase  & Space group (No.) & AFD rotation & Polar/antipolar & \makecell{DFT energy \\ (meV/f.u.)} & \makecell{Model energy \\ (meV/f.u.)} \\
\colrule
1  & $Pm\bar{3}m$  & $Pm\bar{3}m$ (221) &                 &                                                &     0.0  &    0.0  \\
2  & $P4mm$        & $P4mm$       (99)  &                 & FE[1 0 0]                                      &  -202.9  & -208.7  \\
3  & $Amm2$        & $Amm2$       (38)  &                 & FE[1 1 0]                                      &  -232.8  & -235.9  \\
4  & $R3m$         & $R3m$        (160) &                 & FE[1 1 1]                                      &  -260.7  & -263.9  \\
5  & $Pmma$-X      & $Pmma$       (51)  &                 & \makecell{ AFE[1 0 0] \\ (q=[0, 0, 1/2])    }  &   -55.1  &  -56.2  \\
6  & $Cmcm$-AFE    & $Cmcm$       (63)  &                 & \makecell{ AFE[1 1 0] \\ (q=[0, 0, 1/2])    }  &   -63.4  &  -64.0  \\
7  & $Cmmm$        & $Cmmm$       (65)  &                 & \makecell{ AFE[1 0 0] \\ (q=[1/2, 1/2, 0])  }  &  -138.0  & -141.5  \\
8  & $Pmma$-M      & $Pmma$       (51)  &                 & \makecell{ AFE[1 1 0] \\ (q=[1/2, 1/2, 0])  }  &  -122.6  & -129.4  \\
9  & $P4/nmm$      & $P4/nmm$     (129) &                 & \makecell{ AFE[0 0 1] \\ (q=[1/2, 1/2, 0])  }  &    -2.0  &   -3.3  \\
10 & $I4/mmm$-AFE  & $I4/mmm$     (139) &                 & \makecell{ AFE[1 0 0] \\ (q=[1/2, 1/2, 1/2])}  &   -48.5  &  -50.1  \\
11 & $R\bar{3}m$   & $R\bar{3}m$  (166) &                 & \makecell{ AFE[1 1 1] \\ (q=[1/2, 1/2, 1/2])}  &   -66.1  &  -68.6  \\
12 & $P4/mbm$      & $P4/mbm$     (127) & $a^0a^0c^+$     &                                                &  -170.8  & -179.2  \\
13 & $I4/mcm$      & $I4/mcm$     (140) & $a^0a^0c^-$     &                                                &  -192.9  & -192.1  \\
14 & $I4/mmm$-AFD  & $I4/mmm$     (139) & $a^+a^+c^0$     &                                                &  -174.6  & -186.4  \\
15 & $Cmcm$-AFD    & $Cmcm$       (63)  & $a^+b^-c^0$     &                                                &  -230.5  & -229.3  \\
16 & $Imma$        & $Imma$       (74)  & $a^-a^-c^0$     &                                                &  -238.1  & -239.8  \\
17 & $Im\bar{3}$   & $Im\bar{3}$  (204) & $a^+a^+a^+$     &                                                &  -172.4  & -181.7  \\
18 & $P4_2/nmc$    & $P4_2/nmc$   (137) & $a^+a^+c^-$     &                                                &  -227.1  & -226.9  \\
19 & $Pnma$        & $Pnma$       (62)  & $a^-a^-c^+$     &                                                &  -249.4  & -249.4  \\
20 & $R\bar{3}c$   & $R\bar{3}c$  (167) & $a^-a^-a^-$     &                                                &  -222.5  & -227.2  \\
21 & $P4bm$        & $P4bm$       (100) & $a^0a^0c^+$     & \makecell{ FE[0 0 1]                        }  &  -224.9  & -221.6  \\
22 & $I4cm$        & $I4cm$       (108) & $a^0a^0c^-$     & \makecell{ FE[0 0 1]                        }  &  -231.8  & -225.8  \\
23 & $R3$          & $R3$         (146) & $a^+a^+a^+$     & \makecell{ FE[1 1 1]                        }  &  -271.1  & -272.2  \\
24 & $R3c$         & $R3c$        (161) & $a^-a^-a^-$     & \makecell{ FE[1 1 1]                        }  &  -285.3  & -290.3  \\
25 & $Ima2$-FE     & $Ima2$       (46)  & $a^-a^-c^0$     & \makecell{ FE[1 1 0]                        }  &  -276.5  & -277.0  \\
26 & $Pmna$        & $Pmna$       (53)  & $a^-a^-c^0$     & \makecell{ AFE[1 1 0] \\ (q=[1/2, -1/2, 0]) }  &  -256.6  & -256.2  \\
27 & $Pbam$-II     & $Pbam$       (55)  &                 & \makecell{ AFE[1 1 0] \\ (q=[1/4, -1/4, 0]) }  &  -243.6  & -247.7  \\
28 & $Pbam$-AFE40  & $Pbam$       (55)  & $a^-a^-c^0$     & \makecell{ AFE[1 1 0] \\ (q=[1/4, -1/4, 0]) }  &  -285.9  & -286.3  \\
29 & $Pnam$-AFE80  & $Pnam$       (62)  & $a^-a^-c^{+/-}$ & \makecell{ AFE[1 1 0] \\ (q=[1/4, -1/4, 0]) }  &  -286.4  & -286.8  \\
30 & $Ima2$-FiE    & $Ima2$       (46)  & $a^-a^-c^0$     & \makecell{ AFE[1 1 0] \\ (q=[1/3, -1/3, 0]) }  &  -286.7  & -286.3 
\end{tabular}
\end{ruledtabular}
\end{table*}

Thanks to the powerful descriptive capability of neural networks, it is possible to overcome this challenge by constructing a comprehensive training set covering as many phases as possible.
Then, to prevent over-inflating the training set, an efficient strategy for sampling the potential energy surface is also essential. 
Therefore, in contrast to the commonly used methods of concurrently and automatically sampling the potential energy surface from molecular dynamics \cite{RN756}, we construct our training set for PZO through smart design.
We herein briefly illustrate our strategy for the training set design in Figs. \ref{fig:trainingset}(a, b), and more general considerations are presented elsewhere \cite{MultibinitPaper}.
To build the training set, we first condense different unstable lattice modes in the cubic reference structure either individually or jointly, and obtain a series of stationary phases, which are corresponding to the local minima or saddle points on the potential energy surface.
Then, we make linear interpolations and extrapolations between different phases (including the cubic reference phase and the explored distorted phases), and put the interpolating and extrapolating paths into the training set.
Finally, we also include in the training set noisy configurations, which were obtained by populating phonon modes at various temperatures and imposing random strains around each of those phases (the algorithm is implemented in \textsc{Agate}).
The final training set used for training the PZO model contains 12812 configurations in total.
The energy distribution of the configurations is presented in Fig. \ref{fig:trainingset}(c).

\subsection{Model validations}

\subsubsection{Training set and test set}

\begin{figure*}[b]
\centering
\includegraphics[scale=0.48]{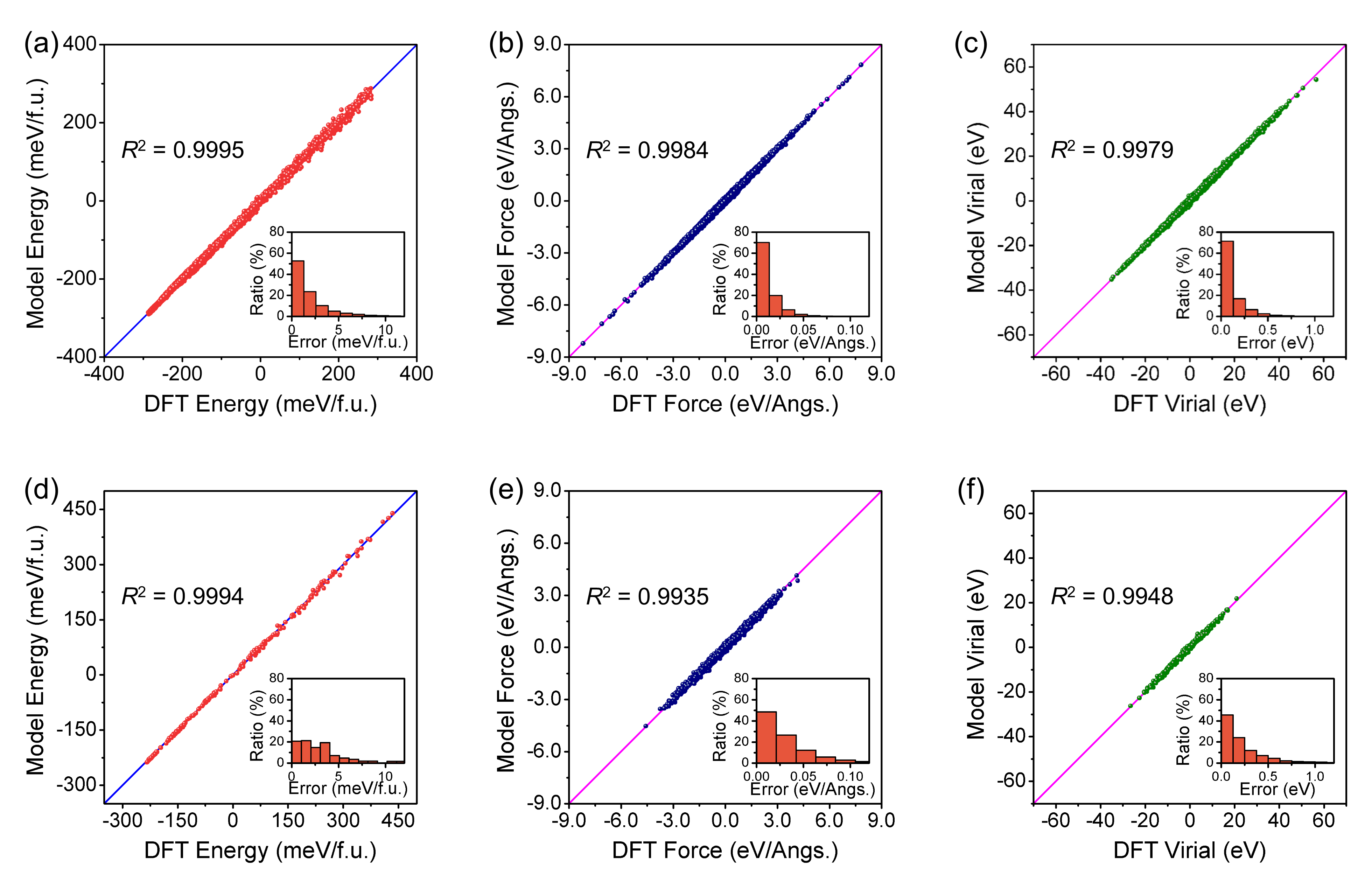}
\caption{Model validations on (a-c) training set and (d-f) test set.
Comparisons of (a, d) energy, (b, e) forces and (c, f) virials between DFT calculations and model predictions.
The insets show the distributions of absolute errors.}
\label{fig:test}
\end{figure*}

We first validate the model by examining its reproducibility on both the training set and a test set.
As shown in Figs. \ref{fig:test}(a-c), our model achieves high-quality fittings on the energy, forces and virials ($R^2 > 0.997$) for the configurations in the training set.
This can be attributed to the powerful descriptive capability of the neural networks.
To ensure that the model does not demonstrate significant overfitting, we construct a test set by randomly selecting 255 configurations from MD trajectories at various temperatures (100 K, 200 K, 300 K, 500 K, 800 K) and feed them back into DFT calculations.
Figs. \ref{fig:test}(d-f) provide comparisons of the energy, forces and virials for the configurations in the test set.
The model also exhibits a high level of reproducibility for the data not in the training set ($R^2 > 0.993$), indicating a satisfactory generalization ability of the model.

\subsubsection{Stationary phases}

\begin{figure*}[b]
\centering
\includegraphics[scale=0.5]{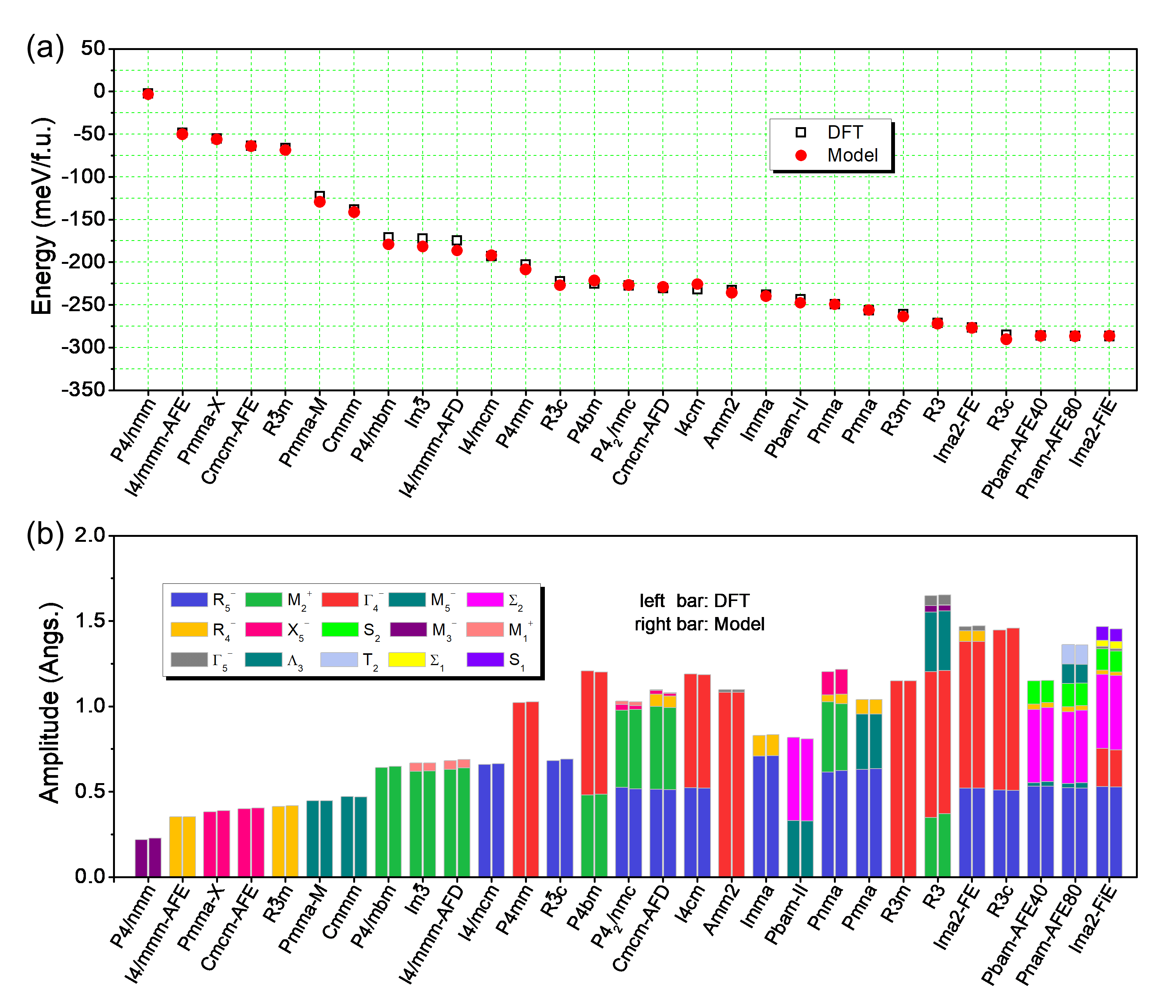}
\caption{Model validations by structural relaxations of different phases, which are corresponding to the stationary points on the potential energy surface.
Comparisons of (a) energy and (b) lattice distortions between DFT calculations and model predictions.
The cubic phase is taken as the energy reference.}
\label{fig:relax}
\end{figure*}

We subsequently validate the model by performing structural relaxations of various stationary phases, to ensure that the corresponding stationary points on the potential energy surface are accurately described.
Due to the generality of our training set, the trained model effectively captures no less than 30 different phases.
Not only the energy [Fig. \ref{fig:relax}(a) and Table \ref{tab:phases}], but also the lattice parameters (Table \ref{tab:lattparam} in Appendix) and atomic distortions [Fig. \ref{fig:relax}(b)] obtained from the model-based structural relaxations are highly consistent with the results obtained from the first-principles relaxations.
In particular, the recent reported $Pnam$-AFE80 and $Ima2$-FiE phases are successfully captured.
This demonstrates that the model is quite efficient in describing the energy landscape.

Nevertheless, there are a few imperfections in the model, particularly the over stabilization of the $R3c$ phase by about 5.0 meV/f.u., leading to a state that exhibits 3 -- 4 meV/f.u. below the $Pbam$-AFE40, $Pnam$-AFE80 and $Ima2$-FiE phases, which should supposedly be above them.
Although this changes the prediction of ground state at zero Kelvin, it does not significantly impact the simulation of the finite-temperature properties, as will be discussed later.
In fact, the energies of these competing low-energy phases are too close to be distinguished within the accuracy of the model and should be considered as nearly degenerated.
We notice that at the DFT level, the exact energy order of these phases already strongly depends on the employed exchange-correlation functional \cite{RN308, PZO_30GS_Ima2}.

\subsubsection{Second-order energy derivatives}

\begin{figure*}[htb]
\centering
\includegraphics[scale=0.5]{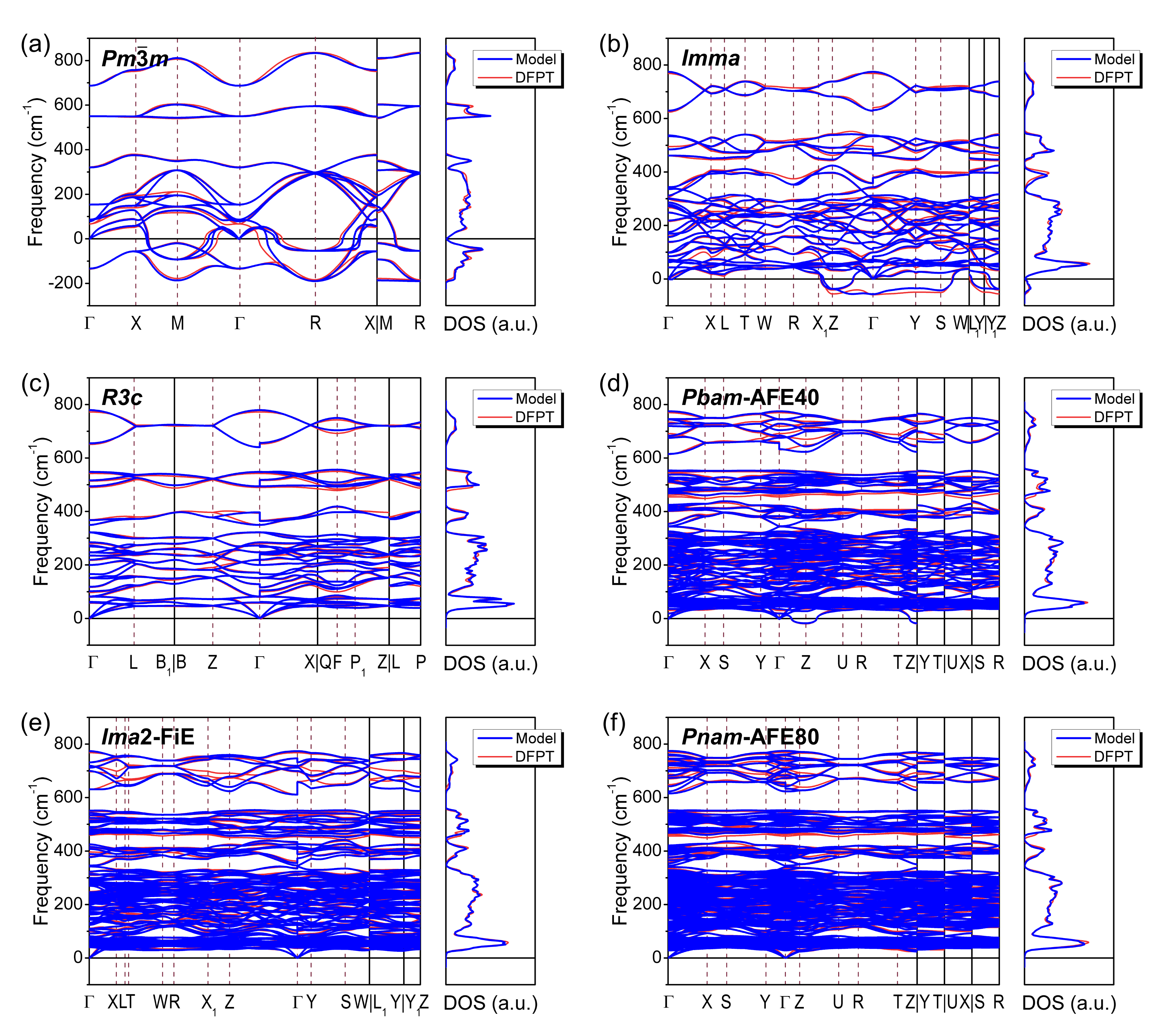}
\caption{Model validations by comparisons of phonon dispersions and phonon density of states calculated from DFPT and the model for different phases of PZO: (a) $Pm\bar{3}m$, (b) $Imma$, (c) $R3c$, (d) $Pbam$-AFE40, (e) $Ima2$-FiE and (f) $Pnam$-AFE80.}
\label{fig:phonon}
\end{figure*}

We further validate the model by calculating the quantities of the second-order energy derivatives, i.e. the phonon frequencies, the elastic and piezoelectric tensors, which are corresponding to the curvatures of the potential energy surface around the stationary points.
The calculations were performed on some important phases, including $Pm\bar{3}m$ (parent cubic phase), $Pbam$-AFE40 (conventional antiferroelectric phase), $Pnam$-AFE80 (newly discovered antiferroelectric phase), $Ima2$-FiE (ferrielectric phase, the DFT ground state), $R3c$ (electric-field induced ferroelectric phase), and $Imma$ phase (the common parent phase of $Pbam$-AFE40, $Pnam$-AFE80, $Ima2$-FiE, and $R3c$ phases).
As illustrated in Fig. \ref{fig:phonon}, and Appendix Tables \ref{tab:ela} and \ref{tab:piezo}, the model provides quite good agreements with the DFPT calculations on these second-order energy derivative quantities.

From Fig. \ref{fig:phonon}, we see that the high-symmetry $Pm\bar{3}m$ and $Imma$ phases exhibit unstable phonon branches [Figs. \ref{fig:phonon}(a, b)], highlighting that they are dynamically unstable at zero Kelvin, while the $R3c$, $Ima2$-FiE and $Pnam$-AFE80 phases are free of lattice instability [Figs. \ref{fig:phonon}(c, e, f)].
For the $Pbam$-AFE40, one imaginary frequency at the $Z$ point can still be observed from both DFPT and model-based calculations [Fig. \ref{fig:phonon}(d)], which is in line with the recent discovery by Baker et al. \cite{PZO_80GS_Pnam}.
By condensing this $Z$-point instability, the $Pbam$-AFE40 phase will change to the $Pnam$-AFE80 phase.

One potential concern with the model is the treatment of long-range dipole-dipole interactions.
It is generally believed that the long-range dipole-dipole interactions are crucial for insulators, and ferroelectric-related phenomena are believed to arise from the delicate balance between long-range and short-range interactions \cite{RN515}.
However, for the model constructed in this work, there is no separation of the long-range and short-range interactions, but the total interactions were treated as a whole and truncated up to a cutoff radius (9 \AA, roughly two times of the cubic cell parameter). 
This suggests that long-range effects might not be so important as initially thought, and some of them may have been effectively incorporated into the current description.
It would be interesting to investigate how the treatment of long-range interactions from other approaches can affect the model \cite{RN1014}.

\section{Finite-temperature properties}
\subsection{Phase transitions}

With the model thoroughly validated, it can be applied to investigate finite-temperature properties of PZO.
Our initial objective is to simulate the temperature dependent phase transitions.
Prior to presenting the results from our model, we will briefly review the experimental facts and the outcomes from previous effective models on PZO.

Experimentally, PZO undergoes a phase transition from the low-temperature $Pbam$-AFE40 phase to the high-temperature cubic phase at $T_{\rm C} \approx 505$ K \cite{RN533}. 
In addition, within a narrow temperature region around $T_{\rm C}$, some experimental studies have also identified traces of a potential intermediate phase, while the exact nature of which is still under debate.

Theoretically, various effective models have been developed to simulate the finite-temperature behaviors of PZO.
Gindele et al. \cite{RN962} developed a shell model for the solid solution $\rm Pb(Zr, Ti)O_3$, in which the antiferroelectric transition of pure PZO is successfully described.
The model predicts the transition occurs around 420 K, slightly underestimated as compared to the experimental value.
In a different approach, Mani et al. \cite{RN52} developed an effective Hamiltonian model for PZO, which successfully describes the antiferroelectric phase transition as well as the behaviors under electric field and mechanical pressure. 
However, their model significantly overestimates the transition temperature to 946 K.
Independently, Patel et al. \cite{RN703} reported another effective Hamiltonian model for PZO, in which a bilinear energetic coupling term between Pb  motions and the oxygen octahedra rotations is introduced.
This model well reproduces the low-temperature $Pbam$-AFE40 phase and the high-temperature cubic phase.
Interestingly, between these two phases, the model predicts an intermediate phase, occurring within the temperature range from 650 K to 1300 K.
This intermediate phase is also in $Pbam$ space group and characterized with similar ``$\uparrow \uparrow \downarrow \downarrow$'' Pb displacement pattern as in the low-temperature $Pbam$-AFE40 phase, but without the oxygen octahedra rotations (corresponding to the $Pbam$-II phase in Table \ref{tab:phases}).
In general, all these existing models mainly focus on the low-temperature $Pbam$-AFE40 phase and the antiferroelectric phase transition to the high-temperature cubic phase, with less consideration given to other potential metastable phases.

A prominent advantage of our model is the capability of describing a large number of metastable phases, and some of them exhibit similarly low energies and may potentially compete with each other.
Our particular interest is to investigate how these competing low-energy phases behave at finite temperatures.
Therefore, we performed MD simulations heating up from different homogeneous phases, i.e. $Pbam$-AFE40 ($Pnam$-AFE80), $Ima2$-FiE and $R3c$, respectively, as well as cooling down from the high-temperature cubic phase.
If not otherwise specified, our calculations employ $12 \times 12 \times 12$ repetitions of the five-atom perovskite unit cell as the simulating supercell, which is commensurate with $Pbam$-AFE40, $Pnam$-AFE80, $Ima2$-FiE and $R3c$, so as to avoid the bias arising from incompatibility in structural periodicity.

\subsubsection{Heating}

\begin{figure*}[htb]
\centering
\includegraphics[scale=0.5]{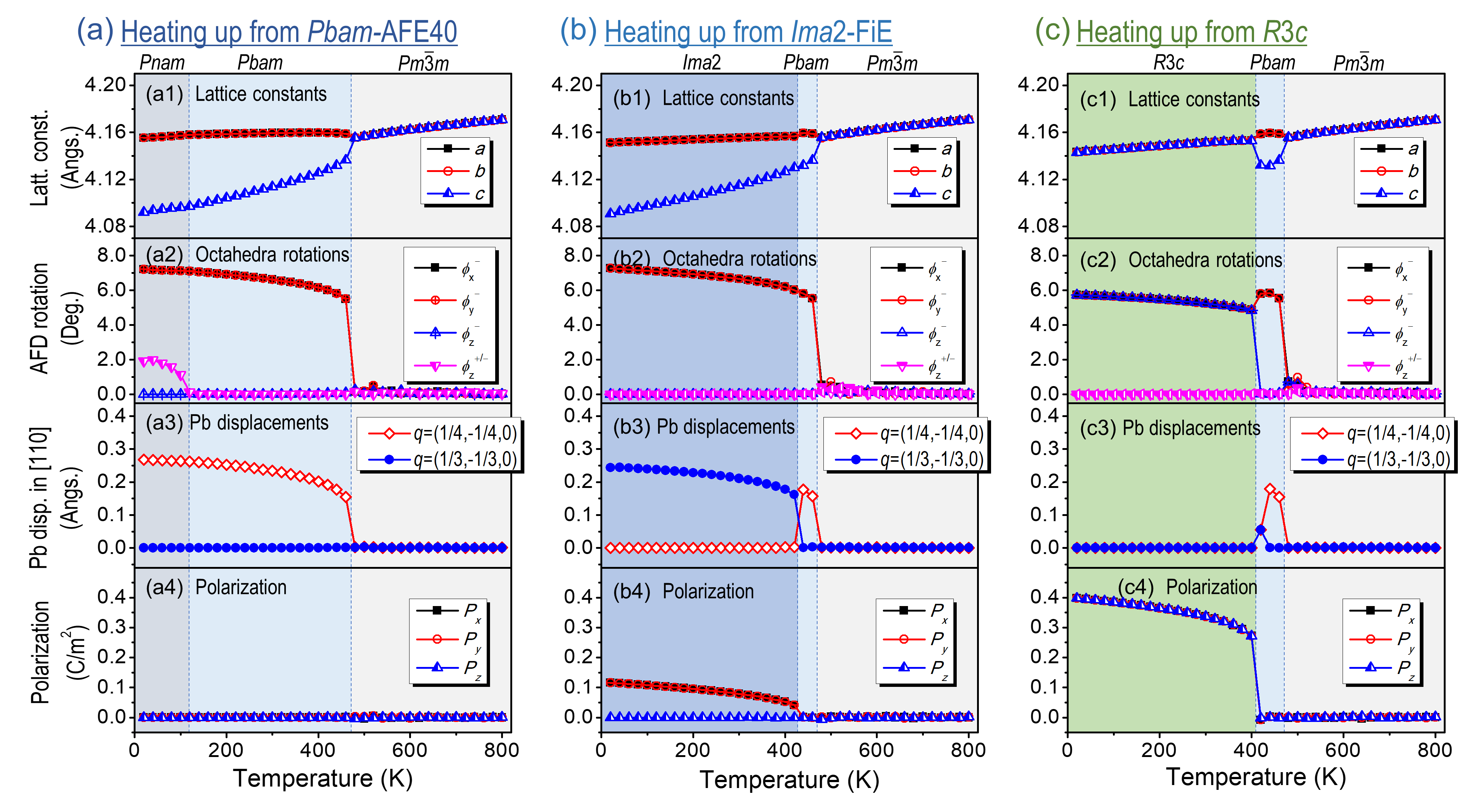}
\caption{Finite-temperature phase transitions during heating up from different homogeneous phases.
The initial structures are in (a) $Pbam$-AFE40, (b) $Ima2$-FiE and (c) $R3c$-FE phases.
For each case, the four panels show the lattice constants, oxygen octahedra rotations, Pb displacements, and polarization as functions of temperature, respectively.
The Pb displacements are first projected to pesudocubic [110] direction and then decomposed by different modulation wave-vectors.
The simulations were performed using $12 \times 12 \times 12$ supercells.
}
\label{fig:heating}
\end{figure*}

When the MD simulation starts heating from the $Pbam$-AFE40 phase, as shown in Fig. \ref{fig:heating}(a), the structure automatically goes to the $Pnam$-AFE80 phase at very low temperature (e.g. 20 K).
This can be evidenced from the non-zero value of $\phi_z^{+/-}$ [Fig. \ref{fig:heating}(a2)].
Here, we use $\phi_z^{+/-}$ as an indicator to distinguish the $Pnam$-AFE80 and $Pbam$-AFE40 phases, which represents the octahedra rotations around the $z$-axis with a complex pattern alternating between in-phase and anti-phase rotations [Fig. \ref{fig:struct}(c)].
It is not surprising that when using $Pbam$-AFE40 as the initial configuration, it immediately transforms into $Pnam$-AFE80, due to condensation of the lattice instability at $Z$-point. 
With increasing temperature upon 120 K, the $Pnam$-AFE80 transforms back to the $Pbam$-AFE40.

When heating up to 480 K, the system changes from the $Pbam$-AFE40 to cubic $Pm\bar{3}m$ phase, along with the vanishing of the octahedra rotations ($a^-a^-c^0$, indicated by $\phi_x^- = \phi_y^- \neq 0$ and $\phi_z^- = 0$) and the antipolar Pb displacements [$\uparrow \uparrow \downarrow \downarrow$, indicated by $q = (1/4, -1/4, 0)$].
We do not observe any clear evidence for possible intermediate phase between $Pbam$-AFE40 and the cubic phase during the heating process.
The $Pbam$-AFE40 to $Pm\bar{3}m$ phase transition temperature observed at 480 K is quite close to the experimental transition temperature $T_{\rm C}^{\rm exp.} \approx 505$ K \cite{RN533}, showing a remarkable improvement on the estimation of the transition temperature compared to previous atomistic simulations \cite{RN52, RN703, RN962}.

When the MD simulations start heating from the $Ima2$-FiE [Fig. \ref{fig:heating}(b)] or $R3c$ [Fig. \ref{fig:heating}(c)] phases, respectively, we found that these phases remain stable at low temperatures, in line with the fact that their phonon dispersions are free of structural instability.
Furthermore, our simulations show that the thermal stability of the $Ima2$-FiE and $R3c$ phases is quite considerable, as they can be maintained to above room-temperature.
This result also implies that there are considerable energy barriers between $Ima2$-FiE and $Pbam$-AFE40 and between $R3c$ and $Pbam$-AFE40.
During heating, the $Ima2$-FiE and $R3c$ first transform into $Pbam$-AFE40 phase around 440 K and 420 K \footnote{Due to the tiny model underestimation of the internal energy of the $R3c$ phase, the transition temperature from $R3c$ to $Pbam$ might be slightly overestimated in our simulations.}, respectively, and then the $Pbam$-AFE40 changes to the cubic $Pm\bar{3}m$ phase around 480 K.
It is very interesting to observe that the phase transition into $Pbam$-AFE40 always occurs in prior to the cubic phase.
Unfortunately, there exists no experimental data for comparison with these simulation predictions so far, probably due to the difficulty of preparing PZO samples in pure $R3c$ or $Ima2$ phases.

\subsubsection{Cooling}

We also make use of the model to simulate phase transitions during cooling process. 
It seems that the simulation of cooling down from the cubic phase is more complicated compared to the simulations of heating up from homogeneous phases.
The complexity arises due to the competition among the nearly degenerated low-energy phases, and the possible emergence of domain walls or phase boundaries during the cooling process.
Nevertheless, the cooling simulation hold the significance of helping to understand the formation of experimental room-temperature phase after high-temperature synthesis or thermal treatment.
In PZO, the most commonly observed room-temperature phase is the $Pbam$-AFE40, but the DFT calculations predicts that some other phases like $Ima2$-FiE would exhibit lower energies.
Why PZO prefers to select the $Pbam$-AFE40 as its room-temperature phase rather than other low-energy phases?
This is a specific question that we want to address through the cooling simulations.

\begin{figure*}[bt]
\centering
\includegraphics[scale=0.5]{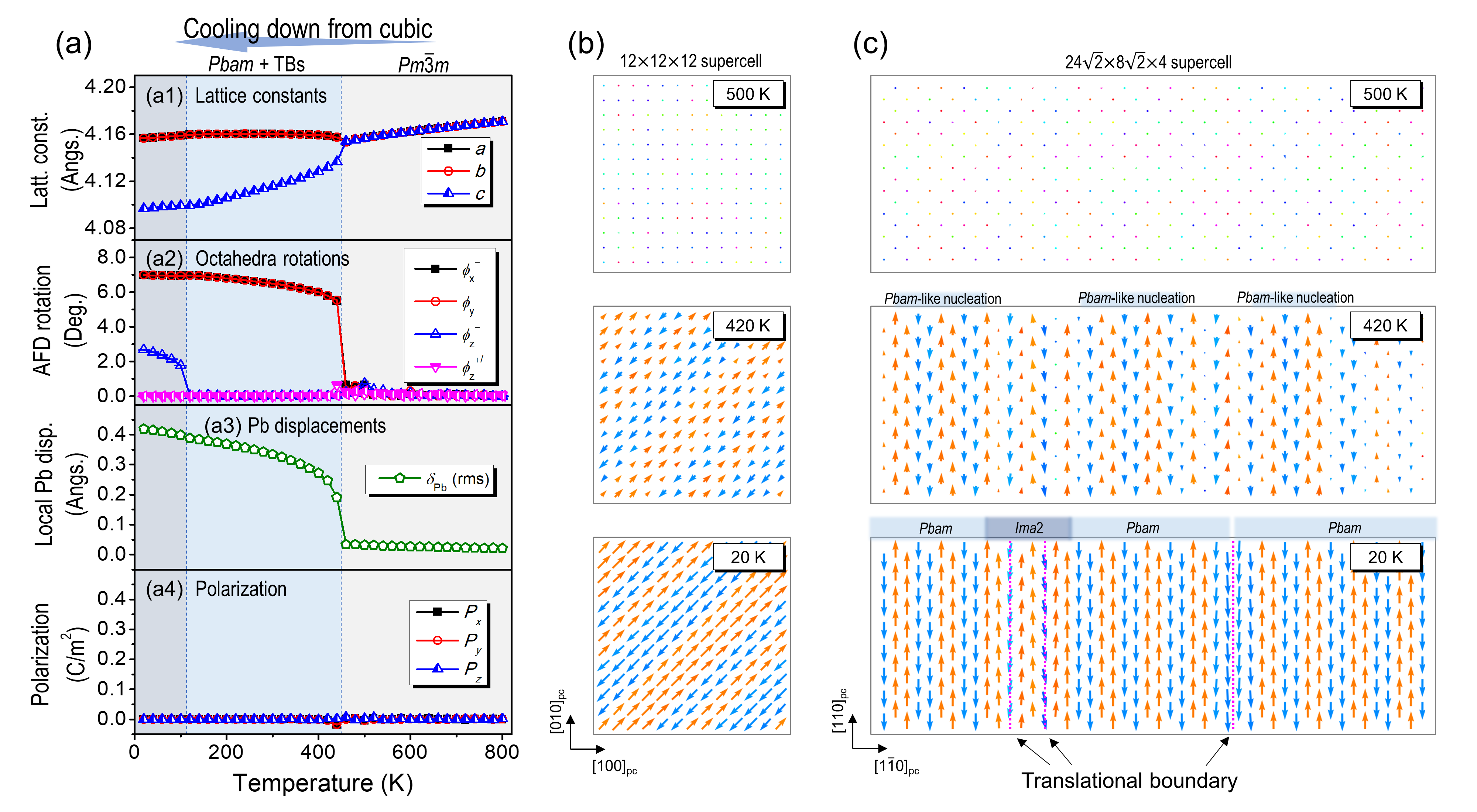}
\caption{Phase transitions and evolution of polarization patterns during cooling down from the cubic phase.
(a) Changes of lattice constants, oxygen octahedra rotations, Pb displacements, and polarization with temperature.
The Pb displacements were characterized by the root-mean-squared displacement with respect to the average position of the 12 surrounding oxygen atoms.
(b, c) Local Pb displacement maps from the average structures at different temperatures during the cooling simulation.
The arrows represent local Pb displacements with respect to the average position of the 12 surrounding oxygen atoms, with colors indicating directions.
The simulations of (a) and (b) were performed using a $12 \times 12 \times 12$ supercells, and the simulations of (c) were performed using a $24\sqrt{2} \times 8\sqrt{2} \times 4$ supercell.}
\label{fig:cooling}
\end{figure*}

Considering the complexity and the inherent stochasticity of the cooling process, we repeated the cooling simulations from the 800 K (cubic phase) to 20 K for several times in supercells of different sizes.
Figs. \ref{fig:cooling}(a, b) present the results from a $12 \times 12 \times 12$ supercell.
The high-symmetry cubic structure go through phase transition below 460 K, along with the appearance of octahedra rotations $a^-a^-c^0$ ($\phi_x^- = \phi_y^- \neq 0$ and $\phi_z^- = 0$) and Pb off-centering displacements, while the system stays as nonpolar during the whole cooling process ($P_x = P_y = P_z =0$).
The transition temperature during cooling is slightly lower than that during heating, indicating the presence of thermal hysteresis, which in line with the fact that the antiferroelectric transition in PZO is of first order.
After cooled down to low temperatures, the final Pb displacement pattern is ``$\uparrow \uparrow \uparrow \uparrow \downarrow \downarrow \uparrow \uparrow \downarrow \downarrow \downarrow \downarrow$'' [Fig. \ref{fig:cooling}(b)], which is different from those in $Pbam$-AFE40 ($\uparrow \uparrow \downarrow \downarrow$), $Ima2$-FiE ($\uparrow \uparrow \downarrow$) or $R3c$ ($\uparrow \uparrow \uparrow \uparrow$), but can be viewed as $Pbam$-AFE40 with two translational boundaries (TBs) \cite{RN853, RN955}.
Besides, an increase in $\phi_z^-$ at the low-temperature end is also observed [Fig. \ref{fig:cooling}(a2)]. 
Upon closer inspection of the low-temperature structure, we find that the non-zero $\phi_z^-$ appears locally in the regions exhibiting the ``$\uparrow \uparrow \uparrow \uparrow$'' and ``$\downarrow \downarrow \downarrow \downarrow$'', indicating a tendency toward $R3c$.
Nevertheless, the emergence of $\phi_z^-$ depends specifically on the Pb displacement pattern and will disappear in simulations using a supercell of a different size.

We repeat the cooling simulation in a $24\sqrt{2} \times 8\sqrt{2} \times 4$ supercell, which allows for a better view of more columns of dipoles [Fig. \ref{fig:cooling}(c)]. 
The changes in lattice parameters, octahedra rotations, and polarization are almost identical to those observed in the simulation using a $12 \times 12 \times 12$ supercell, except for the absence of an increase in $\phi_z^-$ at the low-temperature end.
After cooling down to very low temperature, we find that the prevalence of the $\uparrow \uparrow \downarrow \downarrow$ fragments of Pb displacement pattern is quite noteworthy.
This confirms that the final structure is indeed the $Pbam$-AFE40 phase with translational boundaries.
In this view, the simulation results agree well with the experimental fact that the most commonly observed low-temperature phase of PZO is the $Pbam$-AFE40 phase.

A careful inspection of the evolution of local Pb displacements reveals the detailed process of evolution of the Pb displacement pattern during cooling. 
At the temperature just below $T_{\rm C}$, $Pbam$-like ``$\uparrow \uparrow \downarrow \downarrow$'' domains first nucleate in different regions of PZO [Fig. \ref{fig:cooling}(c), 420 K].
Subsequently, as the temperature decreases, $Pbam$ domains gradually expand and eventually occupy almost the entire crystal. 
Finally, these $Pbam$ domains with $\uparrow \uparrow \downarrow \downarrow$ polar patterns are frozen down to low temperatures [Fig. \ref{fig:cooling}(c), 20 K], while the global transition to other patterns like $Ima2$-FiE ($\uparrow \uparrow \downarrow$), or $R3c$ ($\uparrow \uparrow \uparrow \uparrow$) is kinetically hindered due to the energy barriers between those phases.
The key reason why PZO tends to be in $Pbam$-AFE40 phase at room-temperature is most likely due to the early appearance of $Pbam$-like structures just below $T_{\rm C}$. 
Namely, at the temperature just below $T_{\rm C}$, the $Pbam$-AFE40 holds the advantage in free-energy over other phases such as $Ima2$-FiE or $R3c$. 
Aramberri et al. \cite{PZO_30GS_Ima2} previously made a similar statement, where they estimated the phonon entropy and free-energy using harmonic approximation.
It is encouraging that the statement can now be more directly demonstrated through atomistic simulations.
Our results emphasize the significance of free-energy, rather than the internal energy at zero Kelvin, as a critical factor in determining the actual low-temperature phase in PZO.
 
In addition to the prevalent ``$\uparrow \uparrow \downarrow \downarrow$'' polar patterns, the ``$\uparrow \uparrow \downarrow$'' patterns can also be observed in the region of translational boundaries, which closely resembles the structure of the $Ima2$-FiE phase [Fig. \ref{fig:struct}(b)].
This result is aligned surprisingly well with the recent experimental microscopic observation by Liu et al. \cite{RN853}, who found that in PZO single crystal the $Ima2$-like structures appears at room temperature in the form of translational boundaries.
Because of the random positioning of the initial nucleation of $Pbam$ domains at high temperature, the inevitable outcome is the formation of translational boundaries. 
Based on this, we believe that the translational boundaries are important sources of the $Ima2$-like structures in PZO.

Having these results of heating and cooling, we would also like to discuss the possible impact of the model-predicted energy errors of certain phases on the finite-temperature simulations.
As mentioned before, the model over stabilizes the $R3c$ phase by about 5.0 meV/f.u. (Table \ref{tab:phases}), resulting in the model to incorrectly assign the ground state as the $R3c$ phase.
Despite of this error, the global transitions from $Pbam$-AFE40, $Pnam$-AFE80 or $Ima2$-FiE to $R3c$ have never been observed.
This may be due to the fact that the energy errors are so minor compared to the energy barriers between these phases. 
Another important reason is that the determinant for finite-temperature phase transitions is the free-energy, where the contribution not only involves internal energy but also underscores the crucial role of entropy.
At higher temperatures, the role of entropy becomes even more significant.
Our model can provide phonon density of states highly consistent with DFPT calculations (Fig. \ref{fig:phonon}), indicating that the model can provide a quite good estimation of the lattice vibration entropy.
By these considerations, we speculate that the over stabilization of the $R3c$ phase may lead to the overestimation of the temperature for the $R3c$ to $Pbam$ transition (no experimental result from bulk samples available for comparison), and other than that the impact should be very limited.

\subsection{Double $P$-$E$ hysteresis loop}

Finally, we utilize the model to investigate the electric field induce transitions at finite-temperature.
We first point out that obtaining a typical double $P$-$E$ hysteresis loop in PZO requires specific finite-temperature conditions. 
This is because the ferroelectric $R3c$ phase is dynamically stable [Fig. \ref{fig:phonon}(c)] and possesses a large polarization at zero field [Fig. \ref{fig:heating}(c)], which contradicts the zero remnant polarization for a typical double hysteresis loop.
The optimal temperature for the double $P$-$E$ hysteresis loop should be slightly higher than the $R3c$ to $Pbam$ transition temperature [approximately 420 K according to Fig. \ref{fig:heating}(c), although this value might be overestimated by the model]. 
At this temperature, the polar $R3c$ phase is destabilized under zero field but can be readily induced by the application of an electric field, and the non-polar $Pbam$-AFE40 phase can be restored upon removing the electric field.

\begin{figure*}[b]
\centering
\includegraphics[scale=0.5]{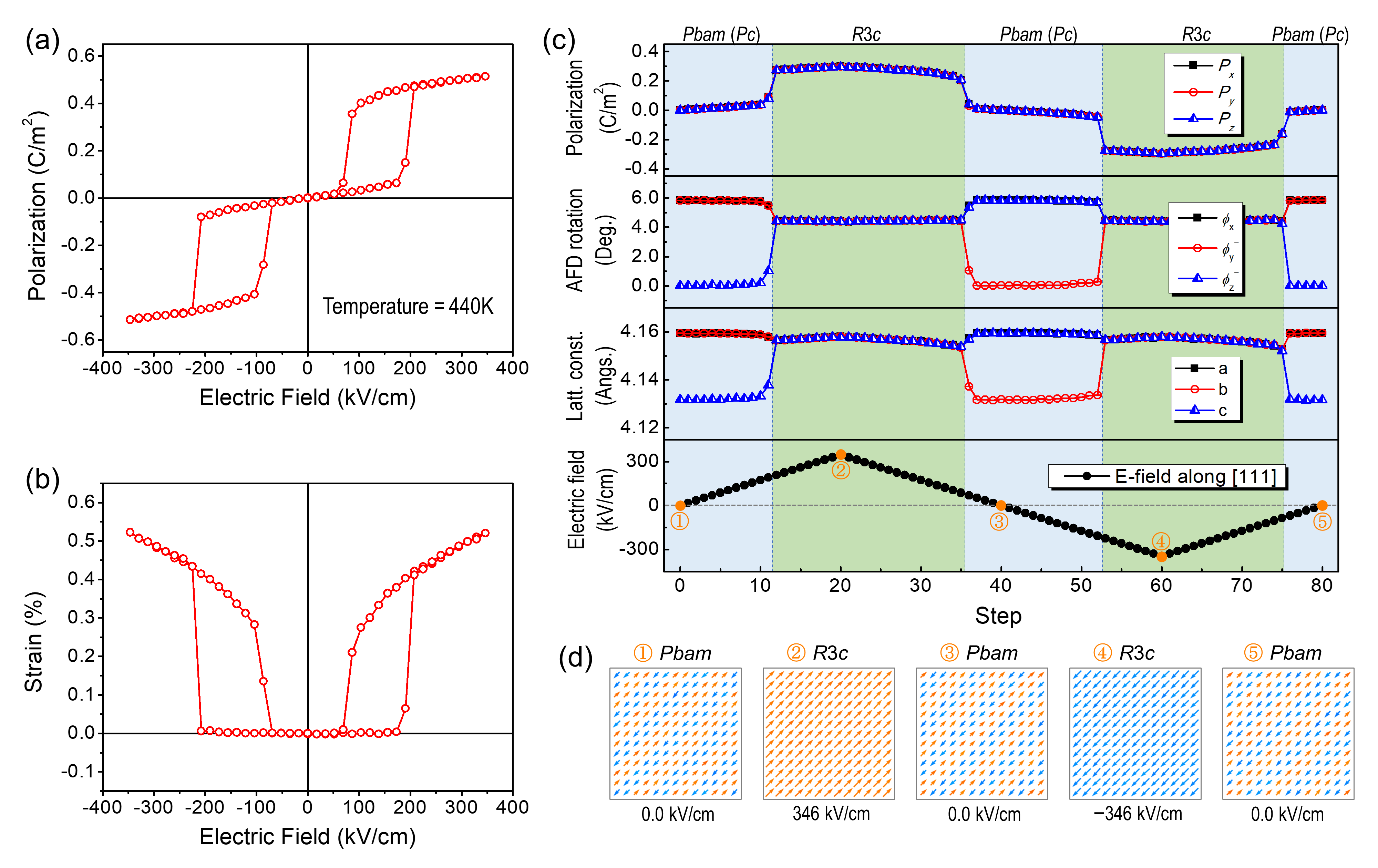}
\caption{Electric-field induced switching between AFE and FE phases.
(a) Polarization hysteresis loop, and (b) strain hysteresis loop.
The polarization was projected into the direction of the electric field, and the strain represents the normal strain component along the direction of the electric field.
(c) Changes of polarization, oxygen octahedra antiphase rotations, lattice constants during the cycling of the electric field.
(d) Local Pb displacement maps from the average structure at different stages of the application of electric field.
The arrows represent local Pb displacements with respect to the average position of the 12 surrounding oxygen atoms, with colors indicating directions.
The simulations were performed using a $12 \times 12 \times 12$ supercell at 440 K, and the electric field is applied along pesudocubic [111] direction.}
\label{fig:PEloop}
\end{figure*}

Figure \ref{fig:PEloop} shows the simulation of double $P$-$E$ hysteresis loop at 440 K.
The electric field was applied along the pseudocubic [111] direction, following a triangle waveform as the shown in Fig. \ref{fig:PEloop}(c).
The simulation provides a double $P$-$E$ hysteresis loop nearly ideal for antiferroelectrics [Fig. \ref{fig:PEloop}(a)].
The maximum polarization $P_{\rm max} = 0.514$ $\rm C/m^2$ (at $E_{\rm max} = 346$ $\rm kV/cm$) is close to the experimental values, and the transition electric fields $E_{\rm AFE-to-FE} \approx 200$ $\rm kV/cm$, $E_{\rm FE-to-AFE} \approx 70$ $\rm kV/cm$ are in the same order of magnitude with the experimental measurements on the PZO films \cite{RN910, RN902, RN905}.
Along with the double $P$-$E$ hysteresis loop, the simulation also provides a sprout-shaped hysteresis loop for the strain, whose shape and the magnitude are also comparable with the experimental measurements on PZO-based materials \cite{RN407, RN803}.

In addition to successfully reproducing the polarization and electric-field induced strain properties, the model-based simulation also reveals the detailed atomic-scale processes of the PZO response to the electric fields.
The changes of Pb displacements and the oxygen octahedra rotation are shown in Figs. \ref{fig:PEloop}(c, d), from which the switching between $Pbam$-AFE40 ($a^-a^-c^0$, $\uparrow \uparrow \downarrow \downarrow$) and $R3c$ ($a^-a^-a^-$, $\uparrow \uparrow \uparrow \uparrow$) during the cycling of electric field is quite clear.
Given that numerous functionalities of PZO, such as dielectric energy storage, electromechanical actuation, electrocaloricity, thermal switching, etc., are all closely linked to electric-field induced AFE-FE transitions, the model also emerges as a potent tool to elucidate the atomistic mechanisms underlying these functionalities.

\section{Conclusion}
We have developed a deep learning interatomic potential for the prototype antiferroelectric perovskite PZO, and demonstrated the capabilities of the model in simulating properties of relatively large systems at finite temperatures and finite electric fields.
The successes of the model include: 
(1) describing a large number of phases, especially the recently discovered $Pnam$-AFE80 and $Ima2$-FiE, and accurately predicting structural and dynamical properties; 
(2) providing $Pbam$ to cubic phase transition temperatures close to the experimental measurement ($T_{\rm C}^{\rm model} \approx 480$ K, $T_{\rm C}^{\rm exp.} \approx 505$ K); 
(3) producing double polarization hysteresis loop that closely resembles the experimental measurement.
The primary weakness of the model is the underestimation of the energy of $R3c$ phase by about 5.0 meV/f.u. 
As a result, the temperature of the transition from $R3c$ to $Pbam$ is likely overestimated.

In this study, the key physical insight provided by the model concerns the formation of $Pbam$ and $Ima2$ structures during cooling. 
Our atomistic simulation shows that at the temperature just below $T_{\rm C}$, the $Pbam$-AFE40 exhibits a free-energy advantage over the other phases.
For this reason, the $Pbam$-like structure appears first, and then frozen down to low temperatures, while the global transition from $Pbam$-AFE40 to $Ima2$-FiE is probably kinetically hindered.
The $Ima2$ structure is more likely to appears locally in the form of translational boundaries.
Due to the stochastic nature of the nucleation of $Pbam$ domains just below $T_{\rm C}$, the formation of translational boundaries at low temperatures is almost inevitable, making them an important source of $Ima2$ structures in PZO.

These results also demonstrate the effectiveness and power of the model-based large-scale atomistic simulations. 
Through these simulations, not only the macroscopic behaviors comparable to experiments can be reproduced, but also the microscale atomistic mechanisms can be accessed.
We prospect that this model will serve as a valuable tool for studying a variety of intriguing phenomena in PZO, including translational boundaries, ferroelastic domain walls, and topological polar structures, as well as functional properties related to energy storage, electric-field induced strain, electrocaloric effect, and thermal switching, etc.

\begin{acknowledgments}
We thank Prof. Gustau Catalan from Catalan Institute of Nanoscience and Nanotechnology and Prof. Bin Xu from Soochow University for helpful discussion.
This work is supported by the European Union’s Horizon 2020 research and innovation program under grant agreement number 964931 (TSAR) and by F.R.S.-FNRS Belgium under PDR grant T.0107.20 (PROMOSPAN). 
H.Z. acknowledges the International Postdoctoral Exchange Fellowship (PC2020060) and the Research IPD-STEMA Program. 
C.G acknowledges financial support from China Scholarship Council.
The authors acknowledge the use of the CECI supercomputer facilities funded by the F.R.S-FNRS (Grant No. 2.5020.1) and of the Tier-1 supercomputer of the Fédération Wallonie-Bruxelles funded by the Walloon Region (Grant No. 1117545). 
\end{acknowledgments}

\newpage
\appendix
\setcounter{table}{0}
\setcounter{figure}{0}
\renewcommand{\thetable}{A\Roman{table}}
\renewcommand{\thefigure}{A\arabic{figure}}
\renewcommand{\thesubsection}{A\arabic{subsection}}

\subsection*{Appendix: Lattice parameters, elastic and piezoelectric tensors of different phases of PZO}

Tables \ref{tab:lattparam}, \ref{tab:ela} and \ref{tab:piezo} provide comparisons of lattice parameters, elastic and piezoelectric tensors of various phases obtained from both first-principles and model-based calculations, respectively, serving as validations of the model. 
Since some of the data have not been previously reported, these tables may also serve as valuable references for future research.

\begin{table*}[h]
\scriptsize
\caption{Lattice parameters of different phases of PZO.}
\label{tab:lattparam}
\begin{ruledtabular}
\begin{tabular}{ccccccccccc}
 & \multirow{2}{*}{Phase}  & \multicolumn{3}{c}{DFT} & \multicolumn{3}{c}{Model} & \multicolumn{3}{c}{Error}  \\
\cmidrule(lr){3-5}\cmidrule(lr){6-8}\cmidrule(lr){9-11}
 & & $a$ (\AA) & $b$ (\AA) & $c$ (\AA) & $a$ (\AA) & $b$ (\AA) & $c$ (\AA) & $a$ (\%) & $b$ (\%) & $c$ (\%) \\
\colrule
1  & $Pm\bar{3}m$  &  4.1402 &   4.1402 &   4.1402 &   4.1401 &   4.1401 &   4.1401 &  -0.001 &  -0.001 &  -0.001 \\
2  & $P4mm$        &  4.1121 &   4.1121 &   4.2579 &   4.1102 &   4.1102 &   4.2630 &  -0.046 &  -0.046 &   0.121 \\
3  & $Amm2$        &  4.1129 &   5.8607 &   5.9957 &   4.1129 &   5.8624 &   5.9929 &   0.001 &   0.030 &  -0.046 \\
4  & $R3m$         &  5.8675 &   5.8675 &   7.2985 &   5.8665 &   5.8665 &   7.2998 &  -0.016 &  -0.016 &   0.018 \\
5  & $Pmma$-X      &  8.3166 &   4.1523 &   4.1548 &   8.3178 &   4.1534 &   4.1545 &   0.014 &   0.026 &  -0.007 \\
6  & $Cmcm$-AFE    &  5.8156 &   5.9437 &   8.3170 &   5.8153 &   5.9460 &   8.3162 &  -0.005 &   0.039 &  -0.010 \\
7  & $Cmmm$        &  8.3155 &   8.4144 &   4.0968 &   8.3140 &   8.4158 &   4.0955 &  -0.017 &   0.017 &  -0.031 \\
8  & $Pmma$-M      &  5.8641 &   4.0971 &   5.9770 &   5.8655 &   4.0967 &   5.9766 &   0.024 &  -0.011 &  -0.006 \\
9  & $P4/nmm$      &  5.8636 &   5.8636 &   4.1415 &   5.8652 &   5.8652 &   4.1408 &   0.026 &   0.026 &  -0.018 \\
10 & $I4/mmm$-AFE  &  5.8846 &   5.8846 &   8.3014 &   5.8844 &   5.8844 &   8.3016 &  -0.004 &  -0.004 &   0.002 \\
11 & $R\bar{3}m$   &  5.8652 &   5.8652 &  14.5461 &   5.8647 &   5.8647 &  14.5521 &  -0.009 &  -0.009 &   0.041 \\
12 & $P4/mbm$      &  5.7711 &   5.7711 &   4.1855 &   5.7709 &   5.7709 &   4.1851 &  -0.003 &  -0.003 &  -0.010 \\
13 & $I4/mcm$      &  5.7715 &   5.7715 &   8.3574 &   5.7691 &   5.7691 &   8.3616 &  -0.041 &  -0.041 &   0.050 \\
14 & $I4/mmm$-AFD  &  8.2663 &   8.2663 &   8.1813 &   8.2633 &   8.2633 &   8.1860 &  -0.037 &  -0.037 &   0.058 \\
15 & $Cmcm$-AFD    &  8.1592 &   8.2638 &   8.2636 &   8.1570 &   8.2636 &   8.2650 &  -0.027 &  -0.003 &   0.017 \\
16 & $Imma$        &  5.8233 &   8.1535 &   5.8685 &   5.8208 &   8.1514 &   5.8711 &  -0.044 &  -0.025 &   0.044 \\
17 & $Im\bar{3}$   &  8.2415 &   8.2415 &   8.2415 &   8.2416 &   8.2416 &   8.2416 &   0.002 &   0.002 &   0.002 \\
18 & $P4_2/nmc$    &  8.2091 &   8.2091 &   8.2784 &   8.2085 &   8.2085 &   8.2753 &  -0.007 &  -0.007 &  -0.038 \\
19 & $Pnma$        &  5.8261 &   8.2175 &   5.8076 &   5.8304 &   8.2105 &   5.8093 &   0.075 &  -0.085 &   0.030 \\
20 & $R\bar{3}c$   &  5.8403 &   5.8403 &  14.1540 &   5.8394 &   5.8394 &  14.1557 &  -0.015 &  -0.015 &   0.012 \\
21 & $P4bm$        &  5.7830 &   5.7830 &   4.2434 &   5.7837 &   5.7837 &   4.2416 &   0.011 &   0.011 &  -0.044 \\
22 & $I4cm$        &  5.7810 &   5.7810 &   8.4637 &   5.7818 &   5.7818 &   8.4608 &   0.015 &   0.015 &  -0.035 \\
23 & $R3$          & 11.7182 &  11.7182 &   7.2440 &  11.7173 &  11.7173 &   7.2352 &  -0.008 &  -0.008 &  -0.122 \\
24 & $R3c$         &  5.8412 &   5.8412 &  14.4261 &   5.8413 &   5.8413 &  14.4347 &   0.001 &   0.001 &   0.059 \\
25 & $Ima2$-FE     &  8.1766 &   5.8811 &   5.9050 &   8.1754 &   5.8810 &   5.9065 &  -0.014 &   0.000 &   0.027 \\
26 & $Pmna$        &  8.1279 &   5.8677 &   5.8903 &   8.1272 &   5.8651 &   5.8910 &  -0.009 &  -0.046 &   0.012 \\
27 & $Pbam$-II     & 11.7915 &   5.9623 &   4.1046 &  11.7941 &   5.9627 &   4.1030 &   0.022 &   0.008 &  -0.040 \\
28 & $Pbam$-AFE40  &  5.8742 &  11.7682 &   8.1723 &   5.8760 &  11.7663 &   8.1724 &   0.030 &  -0.016 &   0.002 \\
29 & $Pnam$-AFE80  &  5.8714 &  16.3651 &  11.7577 &   5.8732 &  16.3664 &  11.7556 &   0.032 &   0.008 &  -0.019 \\
30 & $Ima2$-FiE    &  8.1807 &  17.6294 &   5.8618 &   8.1789 &  17.6303 &   5.8640 &  -0.022 &   0.005 &   0.037 
\end{tabular}
\end{ruledtabular}
\end{table*}

\begin{table*}[tb]
\scriptsize
\caption{Elastic stiffness (unit: $\rm GPa$) calculated by DFPT and the model for different phases of PZO.}
\label{tab:ela}
\begin{ruledtabular}
\begin{tabular}{ccc}
 & DFPT & Model \\
\colrule
%  & Elastic Stiffness (GPa) & Elastic Stiffness (GPa) \\
% \cmidrule(lr){2-2} \cmidrule(lr){3-3}
\makecell{$Pm\bar{3}m$}% \\ $x\parallel[100]_{\rm pc}$ \\ $y\parallel[010]_{\rm pc}$ \\ $z\parallel[001]_{\rm pc}$}
&
$\begin{pmatrix}
     325.6 &     91.5 &     91.5 &          &          &          \\
      91.5 &    325.6 &     91.5 &          &          &          \\
      91.5 &     91.5 &    325.6 &          &          &          \\
           &          &          &     64.8 &          &          \\
           &          &          &          &     64.8 &          \\
           &          &          &          &          &     64.8
\end{pmatrix}$
&
$\begin{pmatrix}                                   
     317.9 &     92.3 &     92.3 &          &          &          \\
      92.3 &    317.9 &     92.3 &          &          &          \\
      92.3 &     92.3 &    317.9 &          &          &          \\
           &          &          &     60.5 &          &          \\
           &          &          &          &     60.5 &          \\
           &          &          &          &          &     60.5
\end{pmatrix}$ \\

\makecell{$Imma$}% \\ $x\parallel[1\bar{1}0]_{\rm pc}$ \\ $y\parallel[110]_{\rm pc}$ \\ $z\parallel[001]_{\rm pc}$\\ $y\parallel$ AFD axis}
&
$\begin{pmatrix}
     281.5 &    129.0 &    112.2 &          &          &          \\
     129.0 &    256.1 &     82.3 &          &          &          \\
     112.2 &     82.3 &    285.3 &          &          &          \\
           &          &          &     66.0 &          &          \\
           &          &          &          &     66.5 &          \\
           &          &          &          &          &     67.4
\end{pmatrix}$
&
$\begin{pmatrix}
     279.1 &    133.2 &    109.2 &          &          &          \\
     133.2 &    257.4 &     79.1 &          &          &          \\
     109.2 &     79.1 &    285.2 &          &          &          \\
           &          &          &     67.5 &          &          \\
           &          &          &          &     65.7 &          \\
           &          &          &          &          &     64.1
\end{pmatrix}$ \\

\makecell{$R3c$}% \\ $x\parallel[1\bar{1}0]_{\rm pc}$ \\ $y\parallel[11\bar{2}]_{\rm pc}$ \\ $z\parallel[111]_{\rm pc}$\\ $z\parallel$ AFD axis}
&
$\begin{pmatrix}
     217.5 &    100.7 &     56.5 &    -11.0 &          &          \\
     100.7 &    217.5 &     56.5 &     11.0 &          &          \\
      56.5 &     56.5 &    145.3 &          &          &          \\
     -11.0 &     11.0 &          &     46.0 &          &          \\
           &          &          &          &     46.0 &    -11.0 \\
           &          &          &          &    -11.0 &     58.4
\end{pmatrix}$
&
$\begin{pmatrix}
     218.3 &    104.0 &     61.0 &     -9.3 &          &          \\
     104.0 &    218.3 &     61.0 &      9.3 &          &          \\
      61.0 &     61.0 &    155.7 &          &          &          \\
      -9.3 &      9.3 &          &     49.3 &          &          \\
           &          &          &          &     49.3 &     -9.3 \\
           &          &          &          &     -9.3 &     57.1
\end{pmatrix}$ \\

\makecell{$Pbam$-AFE40}% \\ $x\parallel[1\bar{1}0]_{\rm pc}$ \\ $y\parallel[110]_{\rm pc}$ \\ $z\parallel[001]_{\rm pc}$\\ $y\parallel$ AFD axis}
&
$\begin{pmatrix}
     260.6 &     86.9 &     96.9 &          &          &          \\
      86.9 &    176.7 &     63.2 &          &          &          \\
      96.9 &     63.2 &    231.3 &          &          &          \\
           &          &          &     58.3 &          &          \\
           &          &          &          &     66.8 &          \\
           &          &          &          &          &     73.0
\end{pmatrix}$
&
$\begin{pmatrix}
     248.8 &     81.5 &     90.5 &          &          &          \\
      81.5 &    174.7 &     58.4 &          &          &          \\
      90.5 &     58.4 &    234.4 &          &          &          \\
           &          &          &     57.6 &          &          \\
           &          &          &          &     65.5 &          \\
           &          &          &          &          &     70.9
\end{pmatrix}$ \\

\makecell{$Pnam$-AFE80}% \\ $x\parallel[1\bar{1}0]_{\rm pc}$ \\ $y\parallel[110]_{\rm pc}$ \\ $z\parallel[001]_{\rm pc}$\\ $y\parallel$ AFD axis}
&
$\begin{pmatrix}
     247.3 &     79.7 &    105.4 &          &          &          \\
      79.7 &    173.7 &     70.4 &          &          &          \\
     105.4 &     70.4 &    220.7 &          &          &          \\
           &          &          &     59.9 &          &          \\
           &          &          &          &     66.7 &          \\
           &          &          &          &          &     74.1   
\end{pmatrix}$
&
$\begin{pmatrix}
     237.6 &     75.5 &     99.5 &          &          &          \\
      75.5 &    170.8 &     64.4 &          &          &          \\
      99.5 &     64.4 &    217.2 &          &          &          \\
           &          &          &     59.9 &          &          \\
           &          &          &          &     65.0 &          \\
           &          &          &          &          &     71.5   
\end{pmatrix}$ \\

\makecell{$Ima2$-FiE}% \\ $x\parallel[1\bar{1}0]_{\rm pc}$ \\ $y\parallel[110]_{\rm pc}$ \\ $z\parallel[001]_{\rm pc}$\\ $y\parallel$ AFD axis}
&
$\begin{pmatrix}
     257.1 &     86.2 &    101.5 &          &          &          \\
      86.2 &    183.6 &     72.4 &          &          &          \\
     101.5 &     72.4 &    229.6 &          &          &          \\
           &          &          &     59.2 &          &          \\
           &          &          &          &     68.0 &          \\
           &          &          &          &          &     75.0   
\end{pmatrix}$
&
$\begin{pmatrix}
     247.1 &     79.6 &     96.2 &          &          &          \\
      79.6 &    178.7 &     68.4 &          &          &          \\
      96.2 &     68.4 &    231.7 &          &          &          \\
           &          &          &     60.2 &          &          \\
           &          &          &          &     65.3 &          \\
           &          &          &          &          &     71.6   
\end{pmatrix}$ \\
\end{tabular}
\end{ruledtabular}
\end{table*}

\begin{table*}[tb]
\scriptsize % \tiny
\caption{Piezoelectric tensor ($e$-type, unit: $\rm C/m^2$) calculated by DFPT and the model for the polar $R3c$ and $Ima2$-FiE phases of PZO.}
\label{tab:piezo}
\begin{ruledtabular}
\begin{tabular}{ccc}
 & DFPT & Model \\
\colrule
 %  & Piezoelectric, $e$-type ($\rm C/m^2$) & Piezoelectric, $e$-type ($\rm C/m^2$) \\
 % \cmidrule(lr){2-2}\cmidrule(lr){3-3}
\makecell{$R3c$}% \\ $x\parallel[1\bar{1}0]_{\rm pc}$, $y\parallel[11\bar{2}]_{\rm pc}$ \\ $z\parallel[111]_{\rm pc}$, $z\parallel$ AFD axis}
&
$\begin{pmatrix}
           &          &          &          &      4.2 &      2.1 \\
       2.1 &     -2.1 &          &      4.2 &          &          \\
       3.0 &      3.0 &      3.7 &          &          &         
\end{pmatrix}$
&
$\begin{pmatrix}                                   
           &          &          &          &      4.1 &      2.0 \\
       2.0 &     -2.0 &          &      4.1 &          &          \\
       2.6 &      2.6 &      3.5 &          &          &         
\end{pmatrix}$ \\

\makecell{$Ima2$-FiE}% \\ $x\parallel[1\bar{1}0]_{\rm pc}$ \\ $y\parallel[110]_{\rm pc}$ \\ $z\parallel[001]_{\rm pc}$\\ $y\parallel$ AFD axis}
&
$\begin{pmatrix}
           &          &          &          &          &      2.0 \\
       1.3 &      1.4 &     -0.08&          &      0.0 &          \\
           &          &          &      0.65&          &         
\end{pmatrix}$
&
$\begin{pmatrix}                                   
           &          &          &          &          &      1.6 \\
       1.2 &      1.6 &     -0.15&          &      0.0 &          \\
           &          &          &      0.50&          &         
\end{pmatrix}$ \\
\end{tabular}
\end{ruledtabular}
\end{table*}

\newpage

\bibliography{Reference}
\bibliographystyle{unsrt}

\end{document}